\documentclass[ aps,%
floatfix,%
final,%
notitlepage,%
oneside,%
onecolumn,%
nobibnotes,%
nofootinbib,%
superscriptaddress,%
showpacs,%
]%
{revtex4}

\usepackage{epsfig}
\usepackage{amsfonts}
\usepackage{amsmath}
\usepackage{epsfig}
\usepackage{graphics}

\newcommand{\beq}{\begin{eqnarray}}\newcommand{\eeq}{\end{eqnarray}}
\newcommand{\beqa}{\begin{eqnarray*}}\newcommand{\eeqa}{\end{eqnarray*}}

\begin{document}

\title{ The study of  double vector charmonium meson production at B-factories \\
within light cone formalism.}
\author{V.V. Braguta}
\email{braguta@mail.ru}
\affiliation{Institute for High Energy Physics, Protvino, Russia}

\begin{abstract}
In this paper the processes $e^+ e^- \to J/\Psi J/\Psi,~ J/\Psi \psi',~ \psi' \psi'$ are considered 
in the framework of light cone formalism. An important distinction of this approach in comparison 
to the approaches used in other papers is that relativistic and leading logarithmic 
radiative corrections to the cross section can be resummed within light cone formalism. 
In this paper the effect 
of this resummation is studied. It is shown that this effect is important especially for the 
production of higher charmonium mesons. The predicted cross sections are in agreement with 
the upper bounds set by Belle collaboration. 
\end{abstract}

\pacs{
12.38.-t,  
12.38.Bx,  
13.66.Bc,  
13.25.Gv 
}

\maketitle

\newcommand{\ins}[1]{\underline{#1}}
\newcommand{\subs}[2]{\underline{#2}}
\vspace*{-1.cm}
\section{Introduction}

The measurement of the cross section of the process $e^+ e^- \to J/\Psi \eta_c$ at Belle collaboration \cite{Abe:2002rb} revealed
large discrepancy between the experiment and the leading order NRQCD prediction. Latter measurements of the 
processes of double charmonium production at B-factories 
\cite{Abe:2004ww, Aubert:2005tj} shown that there is disagreement between theory \cite{Braaten:2002fi, Liu:1, Liu:2}
and experiment in other processes. Only in few years it was realized that the contradiction between 
NRQCD prediction and experimental result for the process $e^+ e^- \to J/\Psi \eta_c$ can  be resolved if one takes into account 
radiative corrections \cite{Zhang:2005ch} and relativistic corrections simultaneously \cite{Bodwin:2007ga, HE:07}. 

In addition to  NRQCD \cite{Bodwin:1994jh}, hard exclusive processes can studied within light cone formalism (LCF) 
\cite{Lepage:1980fj, Chernyak:1983ej}. Within LCF the cross section is built as an expansion 
over inverse powers of characteristic energy of the process. There are two very important advantages of LCF in comparison to NRQCD. 
The first one is connected with the following fact: LCF can be applied to study 
production of any meson. For instance, it is possible to study  production light mesons, 
such as $\pi$ mesons, or production heavy mesons, such as charmonium mesons. 
From  NRQCD perspective, this implies that LCF resums infinite 
series of the relativistic corrections to amplitude, which can be very important. 
The second advantage is that LCF resums very important part of QCD radiative corrections -- 
the leading logarithmic radiative corrections to amplitude $\sim \alpha_s \log (Q)$. This 
is very important advantage since at high energies the leading logarithmic corrections  can be even 
more important than the relativistic ones. 

The first attempts to study double charmonium production at B-factories in the framework of LCF were 
done in papers \cite{Ma:2004qf,Bondar:2004sv,Braguta:2005kr}. The main problem of these papers is connected 
with rather poor knowledge of charmonium distribution amplitudes (DA). It should be noted that 
within LCF the calculation of hard exclusive charmonium production cannot be considered reliable 
if one has poor knowledge of DAs. Fortunately, lately charmonium DAs became the object of intensive study 
\cite{Bodwin:2006dm, Ma:2006hc, Braguta:2006wr, Braguta:2007fh, Braguta:2007tq, Choi:2007ze, Feldmann:2007id}.
In papers \cite{Braguta:2006wr, Braguta:2007fh, Braguta:2007tq}  the models of DAs for 
$1S$ and $2S$ states charmonium mesons were proposed. If one uses these DAs to calculate the cross sections
of the processes $e^+ e^- \to J/\Psi \eta_c, J/\Psi \eta_c', \psi' \eta_c, 
\psi' \eta_c'$, the agreement with the experiments can be achieved \cite{Braguta_new1}. 
In present paper these models of DAs will be used.

A lesson that can be learnt from the study of the process $e^+ e^- \to J/\Psi \eta_c$ within NRQCD and LCF 
is that the leading order NRQCD predictions for hard exclusive processes cannot be considered as reliable
before the relativistic and radiative corrections are not taken into the account. 
This paper is devoted to the study of the hard exclusive processes $e^+ e^- \to J/\Psi J/\Psi,~ J/\Psi \psi',~ \psi' \psi'$ 
in the framework of LCF. An important distinction of this paper in comparison 
to the papers where these processes were studied earlier \cite{Bodwin:2002kk, Bodwin:2002fk, Luchinsky:2003yh, 
Davier:2006fu, Bodwin:2006yd, Gong:2008ce}
is that within LCF the relativistic and leading logarithmic radiative corrections to the cross section can be resummed.
Thus one can hope that the predictions obtained in this way are more reliable. 

This paper is organized as follows. Next section is devoted to the calculation of 
the cross sections of the processes under study at the leading order approximation of LCF. 
In the third section  $1/s$ corrections to the leading order result will be considered. 
In the last section the result of the calculation will be presented and discussed.

\section{The leading order contribution.}

\begin{figure}[t]
\begin{center}
\includegraphics[scale=0.7]{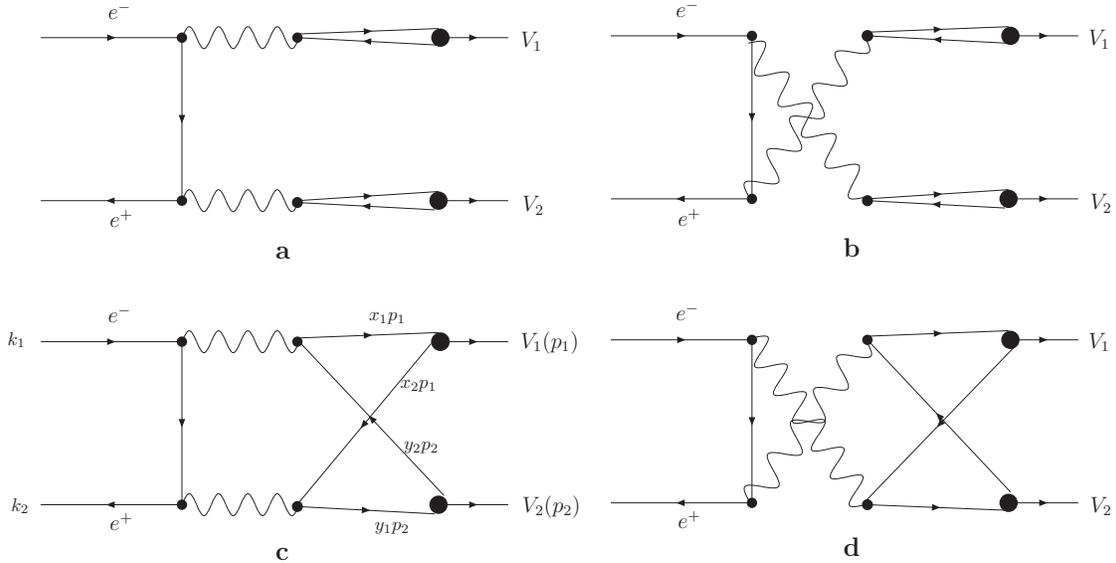} 
\caption{The diagrams that contribute to the process $e^+e^- \to V_1 (p_1) V_2(p_2)$ at the leading order approximation in strong coupling constant.}
\label{tr}
\end{center}
\end{figure}

The diagrams that contribute to the process $e^+e^- \to V_1 (p_1) V_2 (p_2)$ at the leading order approximation in 
$\alpha_s$ are shown in Fig. \ref{tr}. The diagrams shown in Fig. \ref{tr}a, b can be divided into two parts. 
The first part is the annihilation $e^+e^- \to \gamma \gamma$ which is followed by 
the fragmentation of photons into vector mesons $V_1(p_1), V_2(p_2)$. 
Below these diagrams will be referred to as fragmentation diagrams. The second part are the 
diagrams shown in Fig. \ref{tr}c, d
will be referred to as non-fragmentation diagrams. The cross section $\sigma (s)$ of the process $e^+e^- \to V_1 (p_1) V_2 (p_2)$
can be written as the sum 
\beq
\sigma(s) = \sigma_{fr}(s) + \sigma_{int}(s) + \sigma_{nfr}(s),
\eeq
where $\sigma_{fr}(s)$ and $\sigma_{nfr}(s)$ are the contributions due to the fragmentation and non-fragmentation 
diagrams correspondingly, $\sigma_{int}(s)$ is the contribution of the interference between the fragmentation 
and non-fragmentation diagrams.

In this paper double vector charmonium meson production ($V_i=J/\Psi, \psi'$) at B-factories 
will be considered. Commonly, to study charmonium production one uses NRQCD formalism 
\cite{Bodwin:1994jh}. In the framework of NRQCD charmonium mesons are considered as  
nonrelativistic systems with characteristic velocity $v^2 \sim 0.3$ and
the amplitude of charmonium production is the series in the small parameter $v^2$. The study 
of double vector charmonium meson production in $e^+ e^-$ annihilation within NRQCD was carried out in papers 
\cite{Bodwin:2002kk, Bodwin:2002fk, Luchinsky:2003yh, Davier:2006fu, Bodwin:2006yd, Gong:2008ce}.

In this paper light cone formalism \cite{Chernyak:1983ej} will be applied to the study of double vector charmonium 
meson production. Within this formalism the cross section is the series over 
inverse powers of characteristic energy of the process. In particular, at the energy $\sqrt s = 10.6$ GeV
the expansion parameter for the process under consideration is $4  M_V^2/ s \sim 0.4$. 

To begin with let us determine the asymptotic behaviors of  $\sigma_{fr}(s), \sigma_{int}(s), \sigma_{nfr}(s)$ in the 
limit $s \to \infty$. This can be done using the results of paper \cite{Bodwin:2006yd}. In Table \ref{tab1}
the asymptotic behavior of $\sigma_{fr}(s), \sigma_{int}(s), \sigma_{nfr}(s)$ in the limit $s \to \infty$ for 
different polarizations of vector mesons ($\lambda_1, \lambda_2$) are shown. From this table one sees 
that at the leading order approximation in $1/s$ expansion only the fragmentation diagrams with the
polarizations $\lambda_1= \pm 1, \lambda_2 = \mp 1$ contribute. It causes no difficulties to find 
the expression for this contribution
\beq
\frac {d \sigma^{fr}_{\pm 1, \mp 1}} {d x} = \frac { 16 \pi^3 \alpha^4 q_c^4 f_1^2 f_2^2 \sqrt{ \lambda} }
{s M_1^4 M_2^4 } \biggl (  \frac  {1-r_1-r_2} {(1-x^2)  \lambda + 4 r_1 r_2 } \biggr )^2 
\bigl ( 1 - x^4  \bigr ), 
\label{LO}
\eeq
where $M_1, M_2$ are the masses of vector mesons, $q_c$ is the charge of $c$ quark, 
$x = \cos \theta$, $\theta$ is the angle between the momentums of electron and charmonium meson $V_1$, 
\beq
r_1=\frac {M_1^2} s,~~r_2=\frac {M_2^2} s,~~\lambda = 1 + r_1^2+r_2^2-2 r_1 - 2 r_2 -2 r_1 r_2.
\eeq
The constants $f_1$ and $f_2$ are defined through the matrix element of electromagnetic current $J^{em}_{\mu}$ 
\beq
\langle V_i( p_i, \lambda_i)| J^{em}_{\mu}|0 \rangle = q_c f_i \epsilon^*_{\mu}(\lambda_i).
\label{const1}
\eeq
This constants can be determined from the electronic width of vector meson $V_i$
\beq
\Gamma( V_i \to e^+ e^-) = \frac {4 \pi q_c^2 \alpha^2 f_i^2} {3 M_i^3}.
\label{fi}
\eeq
Formula (\ref{LO}) is valid for the production of different mesons. If two 
identical mesons are produced, this formula must be divided by 2. 
It should be noted here that formula (\ref{LO}) is in agreement with the result 
derived in paper \cite{Bodwin:2006yd}. To get the cross section of the process
under consideration one should sum over all possible polarizations that give 
contribution to the cross section at the leading order approximation. Thus up to the corrections
$O(1/s^2)$ the cross section is
\beq
\frac {d \sigma } {d x} = 2 \frac {d \sigma^{fr}_{\pm 1, \mp 1} } {d x } + O \biggl ( \frac 1 {s^2} \biggr )
\eeq

\begin{table}
$$\begin{array}{|c|c|c|c|}
\hline 
V_1 (\lambda_1, p_1) V_2( \lambda_2, p_2) &  \sigma_{fr}(s)  &  \sigma_{int}(s)  &  \sigma_{nfr}(s)  \\
\hline
\lambda_1=\pm 1 ~~\lambda_2= \mp 1  & \sim 1/s & \sim 1/s^2  & \sim 1/s^3  \\
\hline
\lambda_1=\pm 1 ~~\lambda_2=0  &  \sim 1/{s^2} & \sim 1/s^3  & \sim 1/s^4  \\
\lambda_1=0 ~~\lambda_2=\pm 1  &    &   &   \\
\hline
\lambda_1=\pm 1 ~~\lambda_2=\pm 1  & \sim 1/{s^3} & \sim 1/s^4  & \sim 1/s^5  \\
\hline
\lambda_1=0 ~~\lambda_2= 0  & \sim 1/s^3 & \sim 1/s^3  & \sim 1/s^3  \\
\hline
\end{array}$$
\caption{ The leading behavior of $\sigma_{fr}(s), \sigma_{int}(s), \sigma_{nfr}(s)$ in the limit $s \to \infty$ for the polarization 
of vector mesons $\lambda_1, \lambda_2$.  }
\label{tab1}
\end{table}

In this section the cross section of the process $e^+e^- \to V_1 (p_1) V_2 (p_2)$ has been 
considered at the leading order approximation in $1/s$ expansion. Strictly speaking, 
to get the cross section at the leading order approximation one must expand formula 
(\ref{LO}) in $1/s$ and keep only the first term. However, it turns out that 
the contribution of the fragmentation diagrams to the amplitude and the cross section 
can be calculated exactly. So, to improve the accuracy of the calculation done in 
here, the exact expression for cross section due to the fragmentation diagrams 
(\ref{LO}) will be used. 

In the numerical calculation the following values of the parameters will be used:
$M_{J/\Psi}=3.097~ \mbox{GeV}, \Gamma( J/\Psi \to e^+ e^-)=5.55 \pm 0.14~ \mbox{KeV}, 
M_{\psi'}=3.686~ \mbox{GeV}, \Gamma( J/\Psi \to e^+ e^-)=2.48~ \pm 0.06~ \mbox{KeV}$ \cite{Yao:2006px}.
The results of the calculation are shown in Table \ref{tab2}. 
In the second colomn one can see the cross sections at the leading order approximation of $1/s$ expansion.
In the third colomn the differential cross sections at the leading order approximation are integrated over the region $|\cos \theta|<0.8$.

There are different sources of uncertainty to the values of the cross section at 
the leading order approximation of $1/s$ expansion. The first one is QCD radiative corrections. These 
corrections can be divided into three groups. The first group is the radiative corrections 
due to the exchange of gluons between the quark and antiquark of one charmonium meson. 
Evidently, the same corrections appear as the radiative corrections to the electronic
width of charmonium meson. These corrections are included into 
the values of the constants $f_i$, which can determined from the electronic decay 
width of charmonia (\ref{fi}). So, the uncertainty due to the first group of the radiative 
corrections is reduced to the experimental uncertainty in the electronic decay 
widths of charmonia, which is few percents for $J/\Psi$ and $\psi'$ mesons. 
The next group of the radiative corrections are the corrections due to 
the exchange of hard gluons between quarks or antiquarks of different mesons. 
This type of the corrections can be estimated as $ \alpha_s^2( E ) m_c^2/E^2 \sim 0.4 \%$ \cite{Bodwin:2006yd}, where 
$E=\sqrt s / 2$, $m_c$ is the mass of $c$ quark. The last group of the radiative corrections 
is the corrections due to the exchange of soft gluons between quarks or antiquarks of different mesons, 
which can be estimated as $(m_c v)^4 / E^4 \sim 0.04 \%$ \cite{Bodwin:2006yd}. It is seen that 
the uncertainty due to the radiative corrections of the second and third group 
is very small. So, the main source of uncertainty can be reduced to the 
experimental uncertainty in the electronic decay widths of charmonium.

Within light cone formalism in addition to QCD radiative corrections there are power corrections 
to the leading order approximation of $1/s$ expansion. These corrections 
appear due to the contribution of the fragmentation and non-fragmentation diagrams. In Table \ref{tab2}
the error due to this corrections are estimated as $M^2 /E^2 \sim 40 \%$. To reduce this uncertainty 
let us consider $O(1/s^2)$ corrections to the leading order result.

\section{Next-to-leading order contribution in $1/s$ expansion.}

To calculate the cross section at $O(1/s^2)$ approximation of light cone formalism 
let us look to Table \ref{tab1}. It is seen from this table that there are two 
contributions at this level of accuracy. The first one is due to the fragmentation
diagrams with the following polarizations of the mesons 
$\lambda_1 = \pm 1,~ \lambda_2 = 0$ and $\lambda_1 = 0,~ \lambda_2 = \pm 1$. It causes no difficulties 
to calculate these cross sections
\beq
\frac {d \sigma^{fr}_{ \pm 1, 0 }} {d x} =  \frac { 16 \pi^3 \alpha^4 q_c^4 f_1^2 f_2^2 \sqrt{ \lambda} }
{s M_1^4 M_2^4 }  \frac  {2 r_2 } {((1-x^2)  \lambda + 4 r_1 r_2 )^2 }  
\bigl (  (r_1-r_2+1)^2 x^4 + (6 r_1^2 - 2 r_2^2 + 4 r_2 - 2 ) x^2  + (r_1 +r_2 -1)^2 ), \nonumber \\
\frac {d \sigma^{fr}_{  0, \pm 1 }} {d x} =  \frac { 16 \pi^3 \alpha^4 q_c^4 f_1^2 f_2^2 \sqrt{ \lambda} }
{s M_1^4 M_2^4 }  \frac  {2 r_1 } {((1-x^2)  \lambda + 4 r_1 r_2 )^2 }  
\bigl (  (r_2-r_1+1)^2 x^4 + (6 r_2^2 - 2 r_1^2 + 4 r_1 - 2 ) x^2  + (r_1 +r_2 -1)^2 ).
\label{NLO1}
\eeq
The second contribution arises from the interference between the fragmentation and non-fragmentation diagrams 
with polarization of vector mesons $\lambda_1 = \pm 1,~ \lambda_2 = \mp 1$. It is not difficult to 
find the amplitude of the non-fragmentation diagrams (Fig. 1c,d) for these polarization of vector mesons
using LCF
\beq
M=- \frac {2^9 \pi^2 \alpha^2 q_c^2 g_1(\mu) g_2(\mu)} {3 s^3}  
\bigl [ ( 2 (p_2 k_1)-2 (p_1 k_1) ) (\epsilon_1^* \epsilon_2^*) \bar u(k_2)\hat p_1 u(k_1)+ \nonumber \\ 
+ s (\epsilon_1^* k_1) \bar u(k_2)\hat \epsilon_2^* u(k_1) + s (\epsilon_2^* k_1) \bar u(k_2)\hat \epsilon_1^* u(k_1) )  \bigr ]
A (\cos \theta, \mu), 
\label{ampnlo}
\eeq
where $\bar u(k_2), u(k_1)$ are positron and electron bispinors, 
$\epsilon_1^*, \epsilon_2^*$ are polarization vectors of charmonia. The constants $g_i(\mu)$ are defined 
as follows:
\beq
\langle V_i( p_i, \lambda_i)| \bar C \sigma_{\mu \nu} C |0 \rangle_{\mu} = g_i (\mu) \bigl ( \epsilon^*_{\mu} p_{\nu} - \epsilon^*_{\nu} p_{\mu} \bigr ).
\label{const2}
\eeq
It should be noted that the operator ${\bar C} \sigma_{\alpha \beta}  C$ is not renormalization group invariant. 
For this reason the constant $g_i$ depends on scale as
\beq
g_i (\mu) = \biggl ( \frac {\alpha_s( \mu)} {\alpha_s( \mu_0)} \biggr )^{\frac {4} {3 b_0}} g_i (\mu_0).
\eeq
  
The function $A(x, \mu)$ is defined as
\beq
A(x,  \mu) = \frac 1 8 \int d \xi_1 d \xi_2 \frac { \phi_1 (\xi_1, \mu) \phi_2 (\xi_2,  \mu) } { (1+\xi_1 \xi_2)^2 -(\xi_1+\xi_2)^2 x^2 }
\biggl [ \frac 1 {x_1 y_1 } + \frac 1 {x_2 y_2 } \biggr ],
\eeq
here $x_1, x_2$ are the fractions of momentum carried by quark and antiquark in the first meson, 
 $y_1, y_2$ are the fractions of momentum carried by quark and antiquark in the second meson, 
$\xi_1=x_1-x_2, \xi_2=y_1-y_2$, $\phi_1 (\xi_1, \mu),  \phi_2 (\xi_2, \mu)$ are leading twist light cone
distribution amplitudes of vector charmonium mesons with transverse polarization. 

Now some comments on formula (\ref{ampnlo}) are in order:

{\bf 1.} Formula (\ref{ampnlo}) is the leading twist contribution to 
the amplitude of the diagrams shown in Fig \ref{tr}c,d. For this reason it contains only the distribution 
amplitudes $\phi_1 (\xi_1),  \phi_2 (\xi_2)$ of the leading twist. 

{\bf 2.} It is seen from (\ref{ampnlo}) that the amplitude of the non-fragmentation diagrams 
depends on the distribution amplitudes $\phi_i (\xi_i, \mu)$ of vector mesons. 
If infinitely narrow distribution amplitudes $\phi_i (\xi_i, \mu)= \delta( \xi_i)$ are substituted 
to formula (\ref{ampnlo}), than NRQCD result for the amplitude will be reproduced. 
If real distribution amplitudes $\phi_i (\xi_i, \mu)$ are taken at scale 
$\mu \sim m_c$, than formula (\ref{ampnlo}) will resum the relativistic corrections 
to the cross section up to $O(1/s^3)$ terms. 
To resum  the relativistic and leading logarithmic corrections simultaneously
one must take the distribution amplitudes $\phi_i (\xi_i, \mu)$ and the constants $g_i(\mu)$ at the 
characteristic scale of the process $\mu \sim \sqrt s$. The calculation  
of the cross sections  will be done at scale $\mu=E=\sqrt s /2$.

The calculation of $\sigma^{int}_{\pm 1, \mp 1}$ will be done as follows. For the non-fragmentation diagrams 
the amplitude will be taken in form (\ref{ampnlo}). For the fragmentation diagrams 
the exact expression for the amplitudes  will be taken (see discussion in the previous section).
Then the standard procedure for the calculation of the $\sigma^{int}_{\pm 1, \mp 1}$ will be
applied. Thus one gets the result
\beq
\frac {d \sigma^{int}_{\pm 1, \mp 1}} {d x} =- \frac { 2^8 \pi^3 \alpha^4 q_c^4 f_1 f_2 g_1(E) g_2(E) \sqrt{ \lambda} }
{3 s^2 M_1^2 M_2^2 } \biggl (  \frac  {1- r_1 - r_2} {(1-x^2)  \lambda + 4 r_1 r_2 } \biggr ) 
\bigl ( 1 - x^4  \bigr ) A(x,E), 
\label{NLO2}
\eeq
The total cross section has the form
\beq
\frac {d \sigma } {d x} = 2 \frac {d \sigma^{fr}_{\pm 1, \mp 1} } {d x } + 
2 \frac {d \sigma^{fr}_{\pm 1, 0} } {d x } +
2 \frac {d \sigma^{fr}_{0, \pm 1} } {d x } + 
2 \frac {d \sigma^{int}_{\pm 1, \mp 1} } {d x }
+ O \biggl ( \frac 1 {s^3} \biggr )
\label{nlos}
\eeq

\section{Numerical results and discussion. }

From formulas  (\ref{NLO2}), (\ref{nlos}) one sees that  the cross section 
depends on the function $A(x, \mu)$. This function takes into account internal motion of quark-antiquark pairs 
in mesons inside the hard part of the amplitude. In addition, this function resums the leading 
logarithmic radiative corrections to the amplitude. If one ignores both of these effects the 
function $A(x, \mu)$ equals unity, what corresponds to the leading order approximation of NRQCD. 
It is interesting to study how the relativistic and leading logarithmic radiative corrections can change 
the leading order NRQCD predictions. To do this one needs to know  the distribution amplitudes 
$\phi_i(x, \mu)$ of $1S$ and $2S$ states vector charmonium mesons. These distribution amplitudes 
were studied in papers \cite{Braguta:2006wr, Braguta:2007fh, Braguta:2007tq}. The calculation of the functions 
$A(x, \mu)$ and the cross sections of the processes considered will be done using the models 
of distribution amplitudes proposed in these papers:
\beq
\phi_{1S} (\xi, \mu \sim M_c) &\sim& (1- \xi^2) \mbox{ Exp} \biggl [ - \frac {\beta} {1- \xi^2}  \biggr] \nonumber \\
\phi_{2S} (\xi, \mu \sim M_c) &\sim& (1- \xi^2) ( \alpha + \xi^2 ) \mbox{ Exp} \biggl [ - \frac {\beta} {1- \xi^2}  \biggr],
\label{mod}
\eeq
where $M_c=1.2$ GeV is the QCD sum rules mass parameter.
For $1S$ charmonium state the constant $\beta$ can vary within the interval $3.8 \pm 0.7$. 
For $2S$ charmonium state the constants $\alpha$ and $\beta$ can vary within the intervals $0.03^{+0.32}_{-0.03}$ 
and $2.5^{+3.2}_{-0.8}$ correspondingly. 

\begin{figure}[t]
\begin{center}
\hspace*{-1.cm}
\includegraphics[scale=0.89]{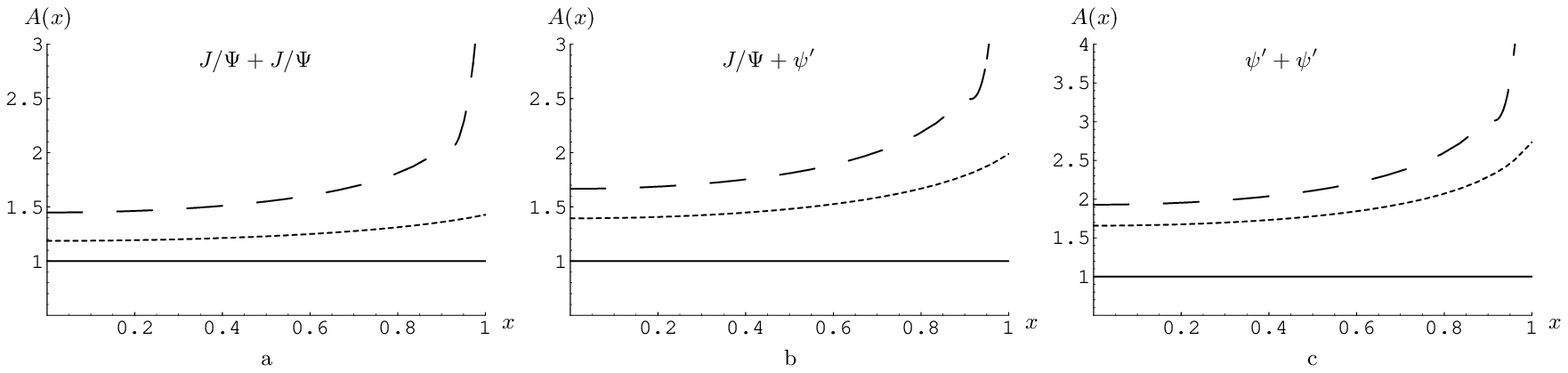} 
\caption{
The plots of the functions $A(x, \mu)$ for the processes $e^+ e^- \to J/\Psi+J/\Psi~(\mbox{{\bf fig.~a}}),
~ J/\Psi+\psi'~(\mbox{{\bf fig.~b}}),~ \psi'+\psi'~(\mbox{{\bf fig.~c}})$. 
Solid lines correspond to the leading order NRQCD predictions for the functions $A(x, \mu)$. 
Small dashed lines represent the functions $A(x,\mu)$ if the relativistic corrections are taken into the account. 
Long dashed lines represent the functions $A(x,\mu)$ if the relativistic corrections and leading 
logarithmic radiative corrections are taken into the account simultaneously. 
}
\label{AA}
\end{center}
\end{figure}

Having models of distribution amplitudes (\ref{mod}), 
it causes no difficulties to calculate the functions $A(x, \mu)$. The plots of the functions $A(x, \mu)$ for 
the processes $e^+ e^- \to J/\Psi+J/\Psi, J/\Psi+\psi', \psi'+\psi'$ are shown in Fig. \ref{AA}a, b, c. 
Solid lines correspond to the leading order NRQCD predictions for the functions $A(x, \mu)$. 
Small dashed lines represent the functions $A(x,\mu)$ if relativistic corrections are taken into the account. 
Long dashed lines represent the functions $A(x,\mu)$ if the relativistic corrections and leading 
logarithmic radiative corrections are are taken into the account simultaneously. 

From Fig. \ref{AA} one sees that the relativistic and leading logarithmic radiative corrections not only can 
change characteristic value of the function $A(x, \mu)$ but they can also considerably modify the  
shape of this function. This statement is especially true for the production of higher 
charmonia  such as $\psi'$ meson, since the relativistic corrections play very important
role in this case.

\begin{figure}[t]
\begin{center}
\hspace*{-1.cm}
\includegraphics[scale=0.89]{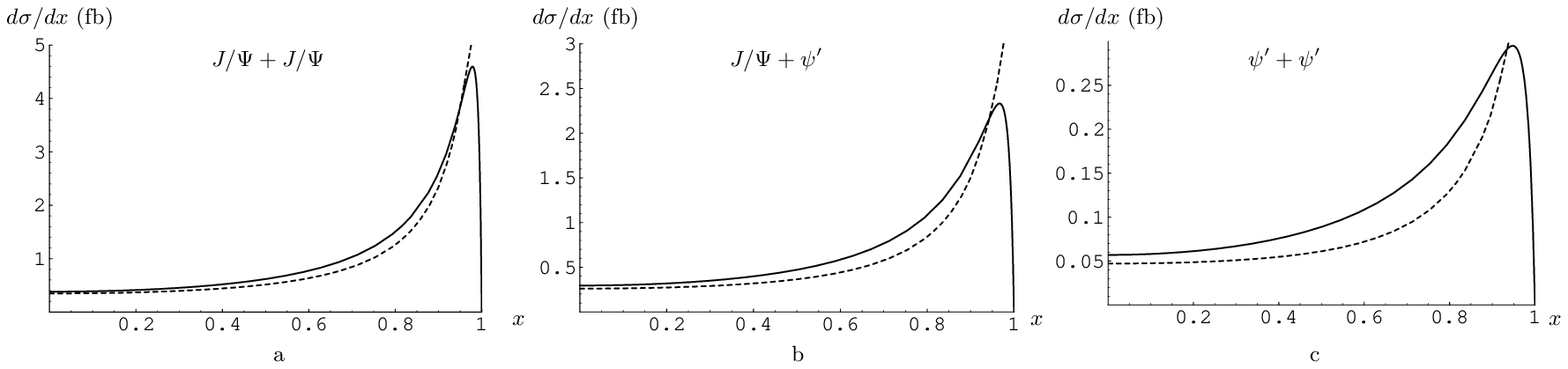} 
\caption{
The plots of the differential cross sections $d \sigma/ d x$ ($x=\cos \theta$) for the processes 
$e^+ e^- \to J/\Psi+J/\Psi~(\mbox{{\bf fig.~a}}),
~ J/\Psi+\psi'~(\mbox{{\bf fig.~b}}),~ \psi'+\psi'~(\mbox{{\bf fig.~c}})$. 
Solid lines correspond to the $O(1/s)$ contributions to the cross sections. 
Small dashed lines represent the cross sections at $O(1/s^2)$ approximation. 
}
\label{sigma}
\end{center}
\end{figure}

Now let us calculate the cross sections of the processes under consideration. To do this 
one needs the values of the constants $g_i(\mu)$ at some scale. Unfortunately, it is rather 
difficult to determine these constants directly from the experiment. The values of these 
constants can be obtained in the framework of NRQCD(see Appendix):
\beq
g_1^2 (M_{J/\Psi}) = 0.144 \pm 0.016~~ \mbox{GeV}^2,~~~~~
g_2^2 (M_{J/\Psi}) = 0.068 \pm 0.022~~ \mbox{GeV}^2. 
\label{const}
\eeq
The plots of the differential cross sections $d \sigma / d x$  for the processes 
$e^+ e^- \to J/\Psi+J/\Psi, J/\Psi+\psi', \psi'+\psi'$ are shown in Fig. \ref{sigma}a, b, c.
The values of the cross sections for the processes under study are shown in Table \ref{tab2}.

There are different sources of uncertainty to the results obtained in this paper. 
The uncertainties in the fragmentation contribution were discussed above. 
The uncertainties in the non-fragmentation contribution can be divided into the following groups:

{\bf 1.} The uncertainty in the models of the distribution amplitudes $\phi_i (x, \mu)$, 
which can be estimated through the variation of the parameters of these models (\ref{mod}).
Thus it is not difficult to show that the error in the cross sections 
due to the uncertainty in the model of the distribution amplitude of $J/\Psi$ meson is not very 
important (about few percents) and it will be ignored further. 
The error due the uncertainty in the model of the distribution amplitude of $\psi'$ meson
is about $10 \%$ of the interference contribution for the process $e^+ e^- \to J/\Psi +\psi'$ and about $20 \%$ 
of the interference contribution for the process $e^+ e^- \to \psi'+ \psi'$.

{\bf 2.} The uncertainty due to the radiative corrections to the non-fragmentation diagrams. 
In the approach applied in this paper the leading logarithmic radiative corrections to the
amplitude have been resummed in the distribution amplitudes.  This fact allows us to estimate 
the rest of the radiative corrections as  $\sim \alpha_s( E ) \sim 20 \%$. It should be 
noted that if one does not resum the leading logarithmic radiative corrections the error 
of the calculation must be estimated as $\sim \alpha_s( E ) \log {s/M_{J/ \Psi}^2 } \sim 50 \%$
instead of  $\sim \alpha_s( E ) \sim 20 \%$  as it was done in paper \cite{Bodwin:2006yd}.  

{\bf 3.} The uncertainty due to the power corrections. This uncertainty is determined 
by the $O(1/s^3)$ terms. One can estimate this source of uncertainty 
as $\sim M^2/E^2 \sim 40 \%$.

{\bf 4.} The uncertainty in the values of constants (\ref{const}). 

Adding all these uncertainties in quadrature one gets the total error of the calculations.

Now it is interesting to compare the results for the cross sections with 
experimental data. The cross sections of the processes considered in this paper 
were measured at Belle collaboration \cite{Abe:2004ww}.  Unfortunately, only the upper bound on these cross sections 
were determined:
\beq
&&\sigma(e^+e^- \to J/\Psi J/\Psi) \times Br_{>2} (J/\Psi ) < 9.1 \mbox{~fb~~~~~~~~~~~~90 \% CL, } \nonumber \\ 
&&\sigma(e^+e^- \to J/\Psi \psi') \times Br_{>2} (\psi' ) < 13.3 \mbox{~fb~~~~~~~~~~~~~~~~90 \% CL, } \nonumber \\ 
&&\sigma(e^+e^- \to J/\Psi \psi') \times Br_{>0} (J/\Psi) < 16.9 \mbox{~fb~~~~~~~~~~~~~90 \% CL, } \nonumber \\ 
&&\sigma(e^+e^- \to \psi' \psi') \times Br_{>0} (\psi') < 5.2 \mbox{~fb~~~~~~~~~~~~~~~~~~~~90 \% CL, }  
\label{exp_res}
\eeq
where $Br_{>2}(V)$ denotes the branching
fraction of $V$ into final states with more than two charged tracks, $Br_{>0}(V)$ is the branching fraction of $V$ into final states
containing charged tracks. Unfortunately, the values of the $Br_{>0,2}(V)$ are unknown. However, one can expect
that the values of the $Br_{>0} (J/\Psi ), Br_{>2} (J/\Psi ), Br_{>0} (\psi')$ are rather close to unity, what allows us to estimate 
$\sigma(e^+e^- \to J/\Psi J/\Psi)<9.1$~fb, $\sigma(e^+e^- \to J/\Psi \psi')<16.9$~fb and $\sigma(e^+e^- \to \psi' \psi') \times Br_{>0} (\psi') < 5.2$~fb.
These estimations are in agreement with the values of the cross sections obtained in this paper.

From the results shown in Fig. \ref{sigma} and in Table \ref{tab2} one sees that 
$O(1/s^2)$ contribution does not change greatly LO results for the process
$e^+ e^- \to J/\Psi J/\Psi$. The smallness of $O(1/s^2)$ contribution for this
process can be explained as follows. At $O(1/s^2)$ approximation there are two contributions 
to the cross sections: the fragmentation diagrams and the interference of the fragmentation and non-fragmentation 
diagrams. These contributions are very near to each other and have different signs. So, due 
to partial cancellation $O(1/s^2)$ contribution to the cross section is suppressed. If one  further considers
the production of higher charmonia, the value of the wave functions at the origin 
for these states are smaller and the total cross section becomes smaller. However, 
due to the relativistic and radiative corrections, collected in the functions $A$, 
the contribution of the non-fragmentation diagrams is enhanced (see Fig. \ref{AA}). 
For this reason $O(1/s^2)$ contribution plays more significant role for the processes
$e^+ e^- \to J/\Psi \psi'$ and it is very important for the process 
$e^+ e^- \to  \psi' \psi'$.

It should be noted here that light cone formalism can be applied to study the production 
of light meson, for instance, $\rho$ mesons. In this case, all formulas derived in this paper 
remain valid. The $O(1/s^2)$ contribution can be estimated 
as $\sim M^2/s$, which is very small value for light mesons. 
For this reason, one can state that the values of the cross sections of double light meson
production obtained in the approximation when only fragmentation diagrams are taken 
into account \cite{Davier:2006fu, Bodwin:2006yd} are rather reliable.

At the end of this section it is interesting to compare the results obtained in this 
paper with the results obtained in other papers devoted to the calculation of the same 
processes. It has already been noted that 
the processes of double vector mesons production were considered in the following papers \cite{Bodwin:2002kk, Bodwin:2002fk, 
Luchinsky:2003yh, Davier:2006fu, Bodwin:2006yd, Gong:2008ce}.  In papers \cite{Luchinsky:2003yh, Davier:2006fu}
 only the contribution arising from the  fragmentation diagrams was considered. 
The results of their calculation are in good agreement with results obtained at 
LO approximation. The papers \cite{Bodwin:2002kk, Bodwin:2002fk, Bodwin:2006yd},
were written by the same group of authors, so it is reasonable to 
consider the results obtained in the last one \cite{Bodwin:2006yd}. 
In this paper the calculation was done at $O(v^2)$ approximation  of NRQCD.
The results obtained in paper \cite{Bodwin:2006yd} are shown in the last two columns 
of Table \ref{tab2}. It is seen that within the error of the calculation this
results are in agreement with that obtained in this paper.

\begin{table}
$$\begin{array}{|c|c|c|c|c|c|c|c|}
\hline 
V_1~ V_2& \sigma^{LO} (\mbox{fb}) & \sigma^{LO}_{|cos \theta | < 0.8} (\mbox{fb}) &   
\sigma^{NLO} (\mbox{fb}) & \sigma^{NLO}_{|cos \theta | < 0.8} (\mbox{fb}) & \sigma^{ \mbox{\cite{Bodwin:2006yd}} } (\mbox{fb}) &
\sigma^{ \mbox{\cite{Bodwin:2006yd}} }_{|cos \theta | < 0.8} (\mbox{fb}) 
& \sigma^{ \mbox{\cite{Gong:2008ce}} } (\mbox{fb})
\\
\hline
J/\Psi~  J/\Psi  & 2.12 \pm 0.85   & 1.02 \pm 0.41 & 2.02 \pm 0.25 & 0.86 \pm 0.17 & 1.69 \pm 0.35 & 0.60 \pm 0.24 & 1.8-2.3 \\
\hline
J/\Psi~  \psi'  & 1.43 \pm 0.57  & 0.77 \pm 0.31 & 1.32 \pm 0.16 & 0.61 \pm 0.16 & 0.95 \pm 0.36 & 0.33 \pm 0.24 & -\\
\hline
\psi' ~  \psi'  & 0.24 \pm 0.10  & 0.14 \pm 0.06 & 0.20 \pm 0.06 & 0.10 \pm 0.05 & 0.11 \pm 0.09 & 0.04 \pm 0.06 & -\\
\hline
\end{array}$$
\caption{ The cross sections of the processes  $e^+ e^- \to J/\Psi  J/\Psi,~ J/\Psi  \psi',~ \psi'  \psi'$. 
The second column contains the cross sections at the leading order approximation of $1/s$ expansion. The third 
column contains the differential cross sections integrated over the region $|\cos \theta|<0.8$.
The values of the cross sections at $O(1/s^2)$ approximation are shown in the forth and fifth columns. 
The sixth and seventh columns contain the results obtained in paper \cite{Bodwin:2006yd}.
The results obtained in paper \cite{Gong:2008ce} are shown in the last column.
}
\label{tab2}
\end{table}

Now let us consider the results obtained in paper \cite{Gong:2008ce}. In this paper the radiative 
corrections to the process $e^+e^- \to J/\Psi J/\Psi$  were calculated. The result of the calculation 
can be written in the following form
\beq
\sigma^{1}=\sigma^0 ( 1+ \frac {\alpha_s} {\pi} K ),
\label{g}
\eeq
where $\sigma^{1}$ is  the cross section with the account of radiative corrections, 
$\sigma^0$ is the cross section without radiative corrections, the factor $K=-11.19$ for the 
pole mass of $c$-quark equals to 1.5 GeV. From this one sees that the radiative corrections
to the cross section are very large, what leads to sizable  reduction of the 
cross section $\sigma^0$. As concerns the results obtained in this paper, it is seen from 
Tab. \ref{tab2} and from Fig. \ref{AA}, \ref{sigma} that the leading logarithmic radiative  corrections 
do not change the cross section greatly. To the first sight, one can think 
that this contradicts to results \cite{Gong:2008ce}. However, there is no contradiction between 
these results. To see this let us consider the results of paper \cite{Gong:2008ce} in more detail. 
One of the input parameter for the calculation of the $\sigma^0$ is the wave function at the origin 
$|R^{J/\Psi}_s (0) |^2$. In paper \cite{Gong:2008ce} this parameter was determined  from the electron 
decay width of $J/\Psi$ through the following formula
\beq
\Gamma_{ee}= \biggl ( 1-\frac {16} 3 \frac {\alpha_s} {\pi} \biggr ) \frac {4 \alpha^2 e_c^2} {M_{J/\Psi}^2} |R^{J/\Psi}_s (0) |^2.
\eeq
It seen that this formula determines the value of $|R^{J/\Psi}_s (0) |^2$ 
taking into the account $\alpha_s$ correction. As the result some part 
of the radiative corrections is present in $\sigma^0$, which according to 
definition is the leading order in $\alpha_s$ quantity. 
So, the authors of this paper separated the whole radiative corrections
into two parts which nearly coincide but have different sign. If one now merges these 
two parts, the following result can be obtained 
\beq
\sigma^{1}=\sigma^0 ( 1+ \frac {32} 3 \frac {\alpha_s} {\pi}  + (-11.19) \frac {\alpha_s} {\pi}  ) =
\sigma^0 ( 1+ (-0.52) \frac {\alpha_s} {\pi} ),
\eeq
what is in agreement with the results obtained in this paper. The value of the 
cross section of the process $e^+e^- \to J/\Psi J/\Psi$ obtained in paper 
\cite{Gong:2008ce} is presented in Table \ref{tab2}. The variation of the cross section is 
due to the variation of the pole mass of $c$ quark $1.4-1.5$ GeV.

\begin{acknowledgments}
The author thanks A.K. Likhoded, A.V. Luchinsky for useful discussion.
This work was partially supported by Russian Foundation of Basic Research under grant 07-02-00417, Russian Education
Ministry grant RNP-2.2.2.3.6646, CRDF grant Y3-P-11-05 and president grant MK-2996.2007.2.

\end{acknowledgments}

\appendix
\section{ The calculation of the constants $g_i(\mu)$. }

To calculate the values of the constants $g_i(\mu)$ (\ref{const2}) one can apply NRQCD formalism. At $O(v^2)$ approximation of NRQCD 
the constants $g_i(\mu)$ and $f_i$  can be written as follows \cite{Bodwin:1994jh, Braaten:1998au}
\beq
\nonumber
f_i^2 &=& \langle V_i(\epsilon) | \chi^+ (\vec {\sigma} \vec {\epsilon }) \varphi | 0  \rangle
\langle 0 | \chi^+ (\vec {\sigma} \vec {\epsilon }) \varphi | V_i(\epsilon)  \rangle 
\times \biggl ( 1-\frac {16} 3 \frac {\alpha_s } {\pi}- \frac 1 3 {\langle v^2 \rangle_i }  \biggr ), \\ 
g_i^2 (\mu) &=& \langle V_i(\epsilon) | \chi^+ (\vec {\sigma} \vec {\epsilon }) \varphi | 0  \rangle
\langle 0 | \chi^+ (\vec {\sigma} \vec {\epsilon }) \varphi | V_i(\epsilon)  \rangle 
\times \biggl ( 1-\frac {16} 3 \frac {\alpha_s } {\pi} - \frac 2 3  \frac {\alpha_s } { \pi} \log {\frac {\mu^2} { m_c^2 } }  
- \frac 2 3  {\langle v^2 \rangle_i } \biggr ),
\eeq
where 
\beq
\langle v^2 \rangle_i =- \frac 1 {m_c^2} 
\frac {\langle 0 | \chi^+ (\vec {\sigma} \vec {\epsilon }) ({\overset {\leftrightarrow} {\bf D} })^2 \varphi | V_i(\epsilon)  \rangle}
 {\langle 0 | \chi^+ (\vec {\sigma} \vec {\epsilon }) \varphi | V_i(\epsilon)  \rangle}.
\eeq
The calculation of the constants $g_i(\mu)$ will be done at scale $\mu=M_{J/\Psi}$.
To diminish the error of the calculation let us consider the ratio $g_i^2(M_{J/\Psi})/f_i^2$. 
At the same level of accuracy it can be written as follows
\beq
\frac { g_i^2(M_{J/\Psi}) } {f_i^2} = \biggl (1-  \frac 2 3 \frac {\alpha_s } 
{ \pi} \log {\frac {M_{J/\Psi}^2} { m_c^2 } } - \frac {\langle v^2 \rangle_i } 3 \biggr ).
\eeq
The values of the constants $g_i (M_{J/\Psi})$ will be calculated with the following set of parameters:
$\alpha_s(M_{J/\Psi})=0.25$, $\langle v^2 \rangle_{J/\Psi}=0.25$ \cite{Bodwin:2006dn}, 
$\langle v^2 \rangle_{\psi'}=0.54$ \cite{Braguta:2007tq}. To estimate the  
error of the calculation one should take into account that within NRQCD the constant is double series 
in relativistic and radiative corrections. At NNLO approximation one has relativistic corrections 
$\sim \langle v^2 \rangle^2$, radiative corrections to the short distance coefficient of the operator
$\langle 0 | \chi^+ (\vec {\sigma} \vec {\epsilon }) \varphi | V_i(\epsilon)  \rangle$ $\sim \alpha_s^2$ and 
radiative corrections to the short distance coefficient of the  operator 
$\langle 0 | \chi^+ ( \vec {\sigma} \vec {\epsilon } ) ({\overset {\leftrightarrow} {\bf D} })^2 \varphi | V_i(\epsilon)  \rangle$
that can be estimated as 
$\sim \alpha_s \langle v^2 \rangle$. Adding all these uncertainties in quadrature one can 
estimate the error of the calculation. Thus one gets
\beq
\nonumber
g_1^2 (M_{J/\Psi}) = 0.144 \pm 0.016~~ \mbox{GeV}^2, \\ 
g_2^2 (M_{J/\Psi}) = 0.068 \pm 0.022~~ \mbox{GeV}^2. 
\eeq

\end{document}